\documentclass[12pt]{article}
\pdfoutput=1
\usepackage[margin=3cm]{geometry}
\usepackage[utf8]{inputenc}
\usepackage{amsmath}
\usepackage{amssymb}
\usepackage{authblk} 
\usepackage{xcolor}
\usepackage{physics}
\usepackage{graphicx} 

\usepackage{hyperref}
\usepackage{varioref}       
\usepackage{cleveref}     

\crefname{table}{table}{tables}
\Crefname{table}{Table}{Tables}
\crefname{figure}{figure}{figures}
\Crefname{figure}{Figure}{Figures}
\Crefname{equation}{Eq.}{Eqs.}

\definecolor{tealblue}{rgb}{0.21, 0.56, 0.63}
\hypersetup{colorlinks=true,allcolors = tealblue,linktocpage=true}

\usepackage{pgf} 

\usepackage[
        backend=bibtex,
        sorting=none,
        style=phys,
        eprint=true,
        doi=false,
        biblabel=brackets
        ]{biblatex}

\bibliography{stringy_banks-zaks.bib}

\newcommand{\calZ}{{\cal{Z}}}

\newcommand{\bQ}{\mathbf{Q}}
\newcommand{\bS}{\mathbf{S}}
\newcommand{\bp}{\mathbf{p}}

\newcommand{\commie}[1]{}

\numberwithin{equation}{section}
\numberwithin{table}{section}

\newenvironment{eqaed}
    {\begin{equation}
    \begin{aligned}
    }
    { 
    \end{aligned}
    \end{equation}
    \ignorespacesafterend
    }

\title{Banks-Zaks Stabilisation of Non-SUSY Strings}

\author{Steven Abel\thanks{s.a.abel@durham.ac.uk}}
\affil{\emph{Institute for Particle Physics Phenomenology, Durham University, Durham DH1 3LE, UK}}
\author{Ivano Basile\thanks{ivano.basile@lmu.de}}
\affil{\emph{Arnold-Sommerfeld Center for Theoretical Physics}\\ \emph{Ludwig Maximilians Universit\"at M\"unchen}\\ \emph{Theresienstraße 37, 80333 M\"unchen, Germany}}
\author{Viktor G. Matyas\thanks{viktor.matyas@liverpool.ac.uk}}
\affil{\emph{Department of Mathematical Sciences, University of Liverpool}\\\emph{Liverpool, L69 7ZL, UK}}

\begin{document}

\maketitle

\begin{abstract}
    
    It appears to be difficult within string theory to obtain genuine scale separation between spacetime and the internal sector. In this paper, we propose a novel mechanism for scale-separated vacua which hinges on stringy effects that are invisible at the level of effective field theory. We show that (meta)stable vacua can form if a super no-scale one-loop potential combines with generic two-loop contributions to the vacuum energy, in a manner analogous to Banks-Zaks fixed points. Weak string coupling and scale separation arise from the accidental smallness of the one-loop term, which receives contributions only from massive states, relative to the two-loop term. We provide a proof of concept of this mechanism in explicit non-supersymmetric heterotic toroidal orbifolds by balancing the complete one-loop contribution against the estimated two-loop term and numerically minimizing the resulting effective potential in a restricted sector of moduli space. We note that this mechanism is generically possible within any fundamental theory that has a one-loop energy that gets contributions only from massive modes.
    
\end{abstract}

\begin{flushright}
LMU-ASC 21/24\\
IPPP/24/77
\end{flushright}

\tableofcontents

\section{Introduction}\label{sec:introduction}
    
    Moduli stabilization in string theory remains an open, highly involved problem. Indeed, the difficulty in achieving stabilization has been crystallized by some results of the so-called swampland program ~\cite{Vafa:2005ui,Ooguri:2006in,vanBeest:2021lhn,Lust:2019zwm} (see, for example, Refs.~\cite{Palti:2019pca,Grana:2021zvf,Agmon:2022thq} for recent reviews) which, among other ideas, suggests that there are inequalities and scaling relations that prevent stable (anti) de Sitter ((A)dS) vacua from forming when supersymmetry is broken or absent. Furthermore, achieving separation between scales is also disfavoured by the AdS distance conjecture \cite{Gautason:2015tig,Gautason:2018gln,Gautason:2019jwq, Lust:2019zwm}. The source of the difficulty lies in the generic link between the mass gap $m$ in an infinite tower of states and the cosmological constant, which is of the form $m \sim |\Lambda|^{\alpha}$ in Planck units, where $\alpha$ is a constant of order one. Scaling relations of this type between UV-sensitive quantities are fine-tuned from the effective field theory viewpoint, and they reflect the UV/IR mixing effects due to gravity. The conjecture of Ref.~\cite{Lust:2019zwm} is that such a relation unavoidably emerges when $\Lambda \to 0$, namely when a family of effective field theories contains parametrically small (nonzero) values of $\Lambda$. If $\alpha \geq \frac{1}{2}$ then there can be no accompanying parametric scale separation, in the sense that the condition  $m \gg \sqrt{\abs{\Lambda}}$ that is required to hide extra dimensions from low-energy observers can not be maintained in the limit  $\Lambda \to 0$. There are no fully controlled constructions of stabilized vacua with parametric scale separation, although ratios of scales that are numerically quite small can be achieved \cite{Demirtas:2021nlu, Demirtas:2021ote}. The closest proposed example is under active scrutiny \cite{DeWolfe:2005uu} and arises with four supercharges. Indeed, several swampland arguments against parametric scale separation and dS vacua in theories with eight or more supercharges have been proposed in the literature \cite{Cribiori:2020use, Cribiori:2023gcy, Cribiori:2023ihv} (see also Ref.~\cite{Cribiori:2024jwq} for a discussion in two dimensions).
    
    Relations between scales of this type almost inevitably arise in any generic non-supersymmetric closed string theory that is expanded around a flat background (regardless of the sign of $\Lambda$) due to modular invariance, which gives a one-loop $\Lambda$ of the form \cite{Dienes:1995pm,Abel:2015oxa,Abel:2021tyt,Abel:2024twz}
    \begin{equation}
\label{eq:lamlam}
        \Lambda_1 ~=~  {\rm Str} (M^2) \frac{{\cal M}^2}{24}~,
        \end{equation}
   where ${\cal M}^2=1/4\pi^2\alpha'$ is the reduced string scale, and where we use a suffix on $\Lambda$ to denote the loop order. Here, the regulated supertrace, which is over the entire infinite tower of masses $M$, yields a finite result that is unavoidably governed by the mass gap. As pointed out in Ref.~\cite{Basile:2024lcz} in the context of Scherk-Schwarz breaking of supersymmetry (SUSY) with radius $R$, this leads to generic difficulties: specifically the issue is that Eq.~\eqref{eq:lamlam} yields the Casimir energy $\Lambda_1 \sim R^{-D}$ where $D$ is the number of noncompact dimensions, which in turn gives $m \sim R^{-1}\sim  |\Lambda_1|^{\frac{1}{D}}$. Thus, it is difficult {\it generically} to separate energy scales hierarchically between, say, $\Lambda$ and the scale of supersymmetry breaking. More precisely, for large $R$ this scenario {\it can} achieve parametric scale separation, but this is only because the dilaton potential is running away exponentially. Similar considerations involving the string coupling (associated with the tower of string excitations in Planck units) have recently been discussed in Ref.~\cite{Montero:2022prj} in the context of the dark-dimension scenario. Thus, at the one-loop level, stabilization of internal moduli {\it is} possible \cite{Ginsparg:1986wr, Angelantonj:2006ut, Fraiman:2023cpa}, but stabilization of the dilaton requires the introduction of fluxes \cite{Mourad:2016xbk}. The resulting Freund-Rubin vacua can be well-controlled and (perturbatively) stable \cite{Basile:2018irz}, but are never parametrically scale separated. 
   
   In this paper, we wish to propose a workaround to these apparent difficulties which is inspired by the analogous situation with renormalization group flow in QCD. Indeed, since the dilaton governs the gauge couplings, the problem of stabilizing the dilaton is, in some sense, a cousin of the problem of finding fixed points in gauge field theories. It is notable that most of the arguments that disfavour a suitable vacuum are based on conclusions drawn from either the classical low-energy effective field theory or perhaps from one-loop corrections or from gaugino condensation or from flux contributions. All of these effects have typically been studied assuming generic leading-order behaviour. However, a similarly generic approach for renormalization in field theory would lead one to conjecture that there could only ever be trivial fixed points where the gauge coupling vanishes, and to conclude that there can be no nontrivial fixed points.  But we know that there {\it are} nontrivial fixed points in QCD, and that these can exist even when the theory is perturbative. Such {\it Banks-Zaks} fixed points occur when the one-loop contribution to the beta function is accidentally small due to a fortuitous choice of the numbers of colours and flavours, while the two-loop contribution is generic.  The latter is then able to balance the former at a fixed point which is still perturbative if the numbers of colours and flavours are chosen to be large. Such perturbative fixed points populate the top of the ``conformal window'' of the QCD phase diagram  \cite{caswell:1974xx, Banks:1981nn}, and they herald the existence of a much larger region in the phase diagram of QCD that includes non-perturbative fixed points.
    
   In an analogous vein for the cosmological constant, we note that there exist theories where the leading ``Casimir energy'' contribution to $\Lambda_1$ coming from Eq.~\eqref{eq:lamlam} vanishes due to an accidental Bose--Fermi degeneracy in the massless sector (and hence all its Kaluza-Klein (KK) excitations) \cite{Abel:2015oxa}. The remaining one-loop contribution to the cosmological constant is exponentially suppressed in such theories, and even its prefactor does not obey the na\"ive scaling relation because the exponentially suppressed contribution comes from massive modes and is an entirely stringy effect.
   Indeed, it is useful to consider a system with Bose-Fermi degeneracy in the massless sector, in which supersymmetry is broken by the Scherk-Schwarz mechanism acting along a single extra compact dimension of radius $R$. In this case the one-loop contribution to the cosmological constant is 
   \begin{equation}\label{eq:one-loop_SS_casimir}
       \Lambda_1(R) ~\approx~ - A (M_1R^{-1})^{\frac{D}{2}} \, e^{- 2\pi M_1 R}~,
   \end{equation} 
   where $M_1$  represents the lowest lying non-zero energy level, which is typically of order $M_1\sim M_s$, and where the constant $A$ reflects the Bose-Fermi non-degeneracy of this energy level.   This is in fact the expected form of $\Lambda_1$ in any theory that has massless Bose--Fermi degeneracy because the leading Casimir energy contribution happens to vanish simply due to the choice of particle content, and the remaining subleading terms are then dominated by the saddle-point approximation of the lowest-lying massive contribution. 
   Importantly, the mass gap $M_1$ in the contributions to $\Lambda_1$ is of the order of the string scale, and thus it is formally divorced from the Kaluza-Klein scale $m\sim R^{-1}$ in the tower (which is also the scale of supersymmetry breaking).

    In such theories, the two-loop contribution to $\Lambda$  becomes crucial. At two-loops the string-frame effective potential takes the following form for closed strings:\footnote{In non-supersymmetric settings where the one-loop contribution $\Lambda_1$ does not vanish, the spacetime theory has dynamical tadpoles, reflecting the fact that the initial configuration is off-shell. In turn, these induce divergences in the two-loop contributions, which should be isolated and removed, leaving the physical two-loop contribution we denote as $\Lambda_2$. Ideally, this procedure would be accompanied by a shift in the background via the Fischler-Susskind mechanism \cite{Callan:1986bc, Callan:1988st, Fischler:1986ci, Fischler:1986tb, Dudas:2004nd, Kitazawa:2008hv, Pius:2014gza, Raucci:2024fnp}.}
\begin{equation}
\label{eq:string-frame_potential}
        V_\text{string} ~=~ \Lambda_1(R) \,+\, \Lambda_2(R) \, e^{2\phi}~,
    \end{equation}
    \vspace{-0.3cm}\\
    where $e^\phi$ is the string coupling. The idea behind our proposed mechanism is then that, because $\Lambda_1 \ll \Lambda_{n>1}$,  stability may be achieved by balancing the accidentally small one-loop term in the cosmological constant against a generically sized two-loop term, in a stringy version of the Banks-Zaks mechanism, which maintains small values of the string coupling and the cosmological constant. Starting with such a stable perturbative minimum, one could then hope to achieve the required final adjustments non-perturbatively in a comprehensive framework. In effect, one would then be balancing stringy one-loop contributions (i.e. terms coming from massive string excitations) against field theoretical two-loop terms (i.e. contributions coming from the massless modes at two-loops). 
    
        We can motivate how this {\it could} happen by invoking the two-loop potential for the single SUSY breaking modulus   of such a compactification, which was discussed in Ref.~\cite{vonGersdorff:2005ce}. At reasonably large modulus values, $R\gtrsim 1$, and neglecting the logarithmic scale dependence (a.k.a. the Coleman-Weinberg potential terms which we shall discuss later) the massless contribution to the two-loop Casimir energy in extended dimensions $D$ take the form     %
    \begin{eqaed}\label{eq:two-loop_casimir}
         \Lambda_2^{\text{massless}} ~\approx~ \frac{B}{R^{D}} 
        ~,
    \end{eqaed}
where $B$ encodes the massless Bose--Fermi nondegeneracy in the two-loop contributions. The expectation for this constant from field theory~\cite{vonGersdorff:2005ce} is that it is of order $g^4/(16\pi^2)^2$ where $g$ is the gauge coupling, times by a multiplicity of order the square of the number of massless states in the theory.
In the Einstein frame in $10$ total dimensions, $V$ is multiplied by $e^{\frac{2 \times 10}{10-2}\phi}$, so that we then have\footnote{With this convention, the dilaton zero-mode (namely the asymptotic string coupling $g_s$) is included in the change of frame. This means that the Planck scale is also rescaled accordingly.}
    \begin{eqaed}\label{eq:einstein-frame_potential}
        V_\text{Einstein} ~=~ e^{\frac{5}{2}\phi} \, \Lambda_1(R) + e^{\frac{9}{2}\phi} \, \Lambda_2(R) ~ .
    \end{eqaed}
    It is easy to show that, neglecting any running in the gauge coupling, the dilaton then stabilizes at
    \begin{eqaed}\label{eq:dilaton_stabilisation}
        e^{2\phi_*} ~=~ - \, \frac{5\Lambda_1(R)}{9\Lambda_2(R)}~ ,
    \end{eqaed}
    \textit{provided} $\Lambda_1(R) \, \Lambda_2(R) < 0$. The Hessian reads
    \begin{eqaed}\label{eq:dilaton_hessian}
       \partial^2_{\phi} V_\text{Einstein}|_{\phi=\phi_*} ~=~ -5\Lambda_1(R) \left(- \, \frac{5\Lambda_1(R)}{9\Lambda_2(R)}\right)^{\frac{5}{4}} \, ,
    \end{eqaed}
    which is positive \textit{provided} $\Lambda_1(R) < 0$, and hence $\Lambda_2(R) > 0$. As we shall verify shortly, this implies that such a minimum must necessarily be an AdS one.
    
    This solution, namely the value of $\phi=\phi_*$, now enters the stabilization equation for the geometric modulus $$\partial_R V_\text{Einstein}(\phi_*,R) ~=~  0~.$$
    Clearly this \textit{does not depend on overall rescalings of $\Lambda_1$ and/or $\Lambda_2$}, and takes the form
\begin{eqaed}\label{eq:geometric_moduli_stabilisation}
       \partial_R \Lambda_1(R) \, - \, \frac{5\Lambda_1(R)}{9\Lambda_2(R)} \, \partial_R \Lambda_2(R) ~=~ 0 ~ ,
    \end{eqaed}
    or, somewhat more compactly,
\begin{eqaed}\label{eq:geometric_moduli_stabilisation_nicer}
       \frac{\partial_R \log \abs{\Lambda_1(R)}}{\partial_R \log \abs{\Lambda_2(R)}} ~=~ \frac{5}{9} 
        ~ .
    \end{eqaed}
    Thus, we arrive at a very general expression for the cosmological constant at the minimum: 
    \begin{eqaed}\label{eq:final_cc}
       \Lambda ~=~ \frac{4}{9} \, \Lambda_1(R_*) \left(- \, \frac{5\Lambda_1(R_*)}{9\Lambda_2(R_*)}\right)^{\frac{5}{4}} ~<~ 0 \, .
    \end{eqaed}
    This is small, due to the accidental hierarchy $\Lambda_1 \ll \Lambda_2$, and negative as anticipated.
    
    Let us now assume that the two-loop contribution (i.e. the constant $B$) is generic and {\it does} have a nonvanishing contribution from the massless modes, as in Eq.~\eqref{eq:two-loop_casimir}, with $B > 0$ as required for the existence of the minimum. Similarly, we assume that the one-loop term in Eq.~\eqref{eq:one-loop_SS_casimir} has $A>0$. Then, taking $D=4$ for concreteness, \Cref{eq:geometric_moduli_stabilisation_nicer} yields
    \begin{eqaed}\label{eq:easy_example_moduli_stabilisation}
       R_* ~=~ \frac{1}{9\pi M_1} ~ ,
    \end{eqaed}
    whereby
    \begin{align}\label{eq:easy_example_cc}
       \Lambda ~&\approx~ - \, 0.024 \, A \, \left(\frac{A}{B}\right)^{\frac{5}{4}} \, M_1^4 \, , \nonumber \\
       g_s & 
       ~\approx~ 0.023 \, \sqrt{\frac{A}{B}} ~ .
    \end{align}
    Thus, in principle (numerically) perturbative AdS minima are possible if the coefficients $A$ and $A/B$ are not too large.  If such a minimum can be achieved, one would then have a (quasi)-stable AdS vacuum that, as mentioned, one could hope to lift with a final stage of minor adjustment. This picture could form the foundation of a (meta)stable phenomenology. The crucial first step then, and the one that will be the focus of this paper, is to show that this schematic picture of stabilization {\it does} actually occur in semi-realistic models.
    
    Although the discussion above gives a broad outline of the mechanism, it is easy to see why showing more rigorously that it actually occurs will require significantly more detail. Indeed, the na\"ive argument above yields a stabilized radius $R_*$, which is relatively close to the string length. Thus, it is not clear {\it a priori} that a simple splitting of contributions into those from massless states and those from a single layer of lowest-lying massive modes is a reasonable approximation. Put differently, one would like to push the minimum to large enough volumes that this approximation becomes valid. Moreover, the net Bose--Fermi number is implicit in $A$ and $B$, and the multiplicity of states increases exponentially with the level; thus $A$ is likely to be greater than $B$. Thus, to properly establish that the mechanism works, it must be performed within an explicit model that has the crucial Bose--Fermi degeneracy of massless modes.  The explicit framework that we shall be using to realize such a scenario in detail is precisely the class of string theories in which SUSY is broken by the Scherk-Schwarz mechanism in Ref.~\cite{Abel:2015oxa}. Ultimately, the picture that will emerge is indeed qualitatively as described.

    We begin the discussion in the next Section by introducing the framework of Scherk-Schwarz compactifications. We describe in detail the one-loop contributions to the vacuum energy, including some limiting regimes. Then we outline some features of the two-loop contribution and our approach to estimate it in light of the difficulties of a full two-loop computation. Then in \Cref{sec:banks-zaks} we discuss the full stabilization mechanism. We close in \Cref{sec:conclusions} with some final remarks on the limitations of our approach and on its motivation as a proof of concept to explore a new corner of the string landscape for scale separation. The main conceptual novelty of this mechanism is that the resulting vacua do not arise from the reduction of a ten-dimensional effective field theory. Rather, the lower-dimensional theory is built directly from the worldsheet and involves stringy effects. Recently, rigid vacua of this type were constructed using other stringy techniques such as asymmetric orbifolds \cite{Baykara:2024tjr, Baykara:2024vss, Angelantonj:2024jtu}. Our approach could in principle be combined with these results perhaps to bypass the most complicated step of our procedure, namely the geometric stabilization, while retaining the crucial features of the mechanism, namely the one-loop Bose--Fermi degeneracy and the correct signs of the one-loop and two-loop terms.
    
\section{Dominant Radiative Contributions}

 In this section, we begin the discussion by collecting all the required radiative contributions in such a framework.  In fact, there are many aspects of radiative contributions that would be expected to be applied to Scherk-Schwarz theories generically, and this will be discussed in the first subsection \ref{subsec:generic}. Indeed, the leading behaviour can be understood in a very geometric fashion, in terms of the propagation properties of the lightest contributing modes. This suffices to understand the general large compactification-radius behaviour described in the introduction.  Even at relatively large radius, in a theory that has the desired massless Bose--Fermi degeneracy, the one-loop cosmological constant will require more detailed knowledge of the massive part of the one-loop partition function. Therefore it is necessary to study a specific model with the desired Banks-Zaks, cancelling leading term, behaviour. We will employ the quasi-realistic  Pati-Salam model introduced in Ref.~\cite{Abel:2015oxa} for this purpose, but will adapt it to a general 2-dimensional toroidal ${\mathbb T}^2$  compactification. This will allow us to include all the relevant compactification moduli, as well as the dilaton. Then in subsection~\ref{subsec:2loop} we collect the generic form of the two-loop contribution for such a compactification, using the results of Ref.~\cite{Abel:2017rch}.

\subsection{The Scherk-Schwarz framework}

The crucial factor that will determine the stabilization is the Scherk-Schwarz action on the different components that appear in the partition function. 
A Scherk-Schwarz action on a theory primarily acts on the space-time charges and the internal momenta of the compact degrees of freedom. In a general closed-string theory, the remaining internal degrees of freedom must then compensate in order to restore modular invariance, and there is a subtle interplay between the two components. 

\noindent\underline{\it The unbroken theory:}

To collect the required ingredients, let us first outline the theory {\it before} any Scherk-Schwarz action is implemented, beginning with the factor in the one-loop partition function coming from the $D+\nu$ large or noncompact bosonic space-time degrees of freedom. Assuming $D$  non-compact dimensions and $\nu$ large compact dimensions, the partition function can be written in the Hamiltonian formalism, that is in terms of the internal momenta of the compactification which depend linearly on the winding and KK numbers ($n^k$ and $m_j$ respectively) as 
\begin{eqnarray}
\mathbf{p}_{L}^{2} & = & p_{L_{i}}G^{ij}p_{L_{j}}\nonumber ~, \\
p_{L_{j}} & = & \frac{1}{\sqrt{2\alpha'}}\left(m_{j}+(B_{jk}+G_{jk})n^{k}\right)~,
\label{eq:PL2}
\end{eqnarray}
and 
\begin{eqnarray}
\mathbf{p}_{R}^{2} & = & p_{R_{i}}G^{ij}p_{R_{j}}\nonumber ~,\\
p_{Rj} & = & \frac{1}{\sqrt{2\alpha'}}\left(m_{j}+(B_{jk}-G_{jk})n^{k}\right)~,
\label{eq:PR2}
\end{eqnarray}
where $G_{ij}$ and $B_{ij}$ are the compact metric and $B$-field.
Henceforth, the $B$-field will be set to zero. 

The partition function factor for the space-time degrees of freedom (including the $\nu$ large compact ones) then takes the form of a partial ``$q$-expansion'':
\begin{equation}
\mathcal{Z}_{D+\nu}(G)~=~\frac{\tau_2^{1-D/2}}{\eta(\tau)^{D+\nu}\bar{\eta}(\bar{\tau})^{D+\nu}}\sum_{\{ m_i,n^j \} \in \{ {\mathbb Z}^{\nu},{\mathbb Z}^{\nu}\} }q^{\alpha'{\bp_{L}^{2}/2}}\bar{q}^{\alpha'{\bp_{R}^{2}/2}},\label{eq:origibos}
\end{equation}
where the world-sheet
parameter is $\tau$, and $q=\exp(2\pi i\tau)$. 

Note that it is convenient to keep $G_{ij}$ dimensionless and write the powers of $\alpha'$ explicitly. Thus in the limit of infinite radius one would Poisson resum to the Lagrangian formalism which transforms the prefactor to ${\alpha'}^{-\nu/2} V_\nu \tau_2^{1-D/2-\nu/2}$ where $V_\nu = \sqrt{\det (\alpha' G_{ij})}$: upon rescaling $Z_{D+\nu}$ by the volume $V_\nu$ one would then recover the partition function for a theory with $D+\nu$ non-compact dimensions as required (and with the correct dimensionality). Therefore it is useful to define a characteristic dimensionful ``compactification radius'' $R$ from the volume as follows:
\begin{align}
\label{eq:vols}
V_\nu ~=~ \sqrt{\det (\alpha' G_{ij})} ~;~~R~=~V^{1/\nu}_\nu   ~, \end{align}
and also to define a normalised and dimensionless compactification metric of determinant 1:
\begin{equation}
    \tilde{G}_{ij} ~=~ \frac{\alpha'}{R^2} G_{ij}  ~.
\end{equation}

In order to maintain modular invariance, the space-time boson factor $\calZ_{D+\nu}$ is accompanied by a factor coming from remaining internal degrees of freedom, which we do not yet need to specify in detail, so that in general  
\begin{equation}
\mathcal{Z} ~=~\mathcal{Z}_{D+\nu} \, \mathcal{Z}_{\text{int}}~,
\end{equation}
where $\mathcal{Z}_{\text{int}}$ is the internal partition
function. To be concrete we can consider cases where ${\mathcal Z}_{\rm int}$ takes the form of a charge lattice with a generic structure that can also be written in the form of a partial $q$-expansion as 
\begin{equation}
{\cal Z}_{\rm int} ~=~\frac{1}{\eta^{26-D-\nu}\bar{\eta}^{14-D-\nu}}  \sum_{\{ \bQ_L, \bQ_R \} }
e^{2\pi i \bQ\cdot \bS}
q^{{\bQ_{L}^{2}/2}}\bar{q}^{{\bQ_{R}^{2}/2}}~,\label{eq:origibos2}
\end{equation}
where in the case of the heterotic string $\bQ_{R}$ and $\bQ_{L}$ are $14-D-\nu$ and $26-D-\nu$ dimensional charge vectors respectively, and the phase prefactor is designed to give the correct space-time spin statistics to the states by virtue of the spin-vector $\bS$. The sum above over charges and phases must, of course, be carefully structured to maintain modular invariance. In practice, this structure is implemented by means of a sum over sectors and a set of GSO projections, expressed in terms of a basis for the internal $(14-D-\nu,26-D-\nu)$-vectors. Hence in order to see the generic effect of the Scherk-Schwarz action, it will be very useful to keep in mind the following alternative formulation of the internal partition function:
\begin{equation}
{\cal Z}_{\rm int}~=~\frac{1}{\eta^{26-D-\nu}\bar{\eta}^{14-D-\nu}} \sum_{\{ \alpha_a, \beta_b\} }  \Omega\left[
\begin{array}{c}
\alpha\\
\beta
\end{array}\right] ~,
\end{equation}
where the sum over the $\beta_b$ basis coefficients implements a projection on the states, while the $\alpha_a$ sum produces the different sectors. The functions $\Omega$ are modular covariant combinations that include phases which are arranged to render the entire ${\cal Z}$ modular invariant. For our discussion, we do not need to display them explicitly as this is rather standard and our aim is to be as general as possible: however, some examples are given in Ref.~\cite{Abel:2015oxa}.\\

\noindent\underline{\it The Scherk-Schwarzed theory:}

Having laid out the general form of the supersymmetric theory, we can now proceed to the effect of a Scherk-Schwarz (SS) action in the compactification. It induces a fractional shift in the KK numbers of gravitini (and various other states, usually but not necessarily including all the gaugini). 
In addition to the shift in KK number, the Scherk-Schwarz mechanism is defined by a simultaneous action on the Cartan charges
of a state. Letting the Scherk-Schwarz action be described by the   $(14-D-\nu,26-D-\nu)$-vector $\mathbf{e}$
(which crucially overlaps the $R-$charge in order to distinguish
fermion from boson), one has (see for example Ref.~\cite{Kiritsis:1997hf}),
\begin{equation}
\mathbf{Q}~\longrightarrow~\mathbf{Q}-n\ \mathbf{e} \qquad ;\qquad \ \  m_{i}~\longrightarrow~ m_{i}+\mathbf{Q}\cdot \mathbf{e} - \frac{n}{2}\ {\bf e}\cdot{\bf e} ~,\label{eq:Wilson}
\end{equation}
where $n=\sum_i n^i$, where  $\mathbf{e}\cdot\mathbf{Q}$ is a fraction that depends on the compactification (e.g. the orbifold), and where the dot product is Lorentzian. The fractions need not always be integer or half integer: for example, in ${\mathbb Z}_N$ orbifolds the KK shifts could be in fractions $1/N$ if $N$ is even or $1/2N$ if $N$ is odd. (Note that one need only incorporate a single SS action,
as the sum of two different actions is just a Wilson line.) However, any non-trivial phases must, of course, add up to real contributions in the final partition function, so the overall effect in ${\cal Z}$ must be carefully orchestrated for consistency.

To see how, let us define the possible fractional KK shifts as \begin{equation}
    {k}_e/N  \,~\stackrel{\rm def}{=}~\,
    \mathbf{e\cdot Q} - \frac{n}{2}\ {\bf e}\cdot{\bf e} \,\,\, {\rm mod}(N) ~,
    \end{equation}
where the sum over sectors can be taken without loss of generality to involve fractional shifts $k_e/N$ where $k_e\in \{ 0\ldots N-1\}$.  The effect of the
Scherk-Schwarz compactification is to then split the spectrum into sectors
that depend on both this shift and on the winding numbers. Following  Ref.~\cite{Aaronson:2016kjm},
the partition function becomes split according to the $k_e$ values, as 
\begin{equation}
\mathcal{Z} ~=~\frac{1}{N}\sum_{\bar{n},{k}_{e}=0}^{N-1}\mathcal{Z}_{D+\nu, \, \bar{n}, {k}_e} \, \mathcal{Z}_{\text{int},\bar{n} ,{k}_{e}}~,
\label{eq:partitionfuncgen}
\end{equation}
where $\bar{n} ~=~\sum_i n_{i}\,\,\,\text{mod}(N)$, where
$\mathcal{Z}_{D+\nu, \, \bar{n}, {k}_e} $ is the bosonic partition function with KK number $m_i$ shifted by $k_e/N$ as in Eq.~\eqref{eq:Wilson} and with sum the over winding restricted to the specified value of $\bar{n}$, 
and where $\mathcal{Z}_{\text{int},\bar{n} ,{k}_{e}}$ is the shifted internal partition
function,
\begin{equation}
\mathcal{Z}_{\text{int},\bar{n},{k}_{e}}~=~\frac{1}{N}\sum_{\alpha,\beta,\beta_{e}}e^{-2\pi i\beta_{e}k_{e}/N}\,\Omega
\left[\begin{array}{c}
\{\alpha,-\bar{n}\}\\
\{\beta,\beta_{e}\}
\end{array}\right]~.
\label{eq:shiftenZint}
\end{equation}
According to Eq.~\eqref{eq:shiftenZint} the SS action now sits inside  ${\cal Z}_{\rm int}$ like a
 new ${\bf e}$ basis vector. In other words the Scherk-Schwarz action entangles the previously separate ${\cal Z}_{D+\nu}$ and ${\cal Z}_{\rm int}$ such that the latter comes in $N$ different sectors labelled by $\bar{n}$ which loops over the number of ${\bf e}$ vectors in the shifted charge. Moreover the KK shift $k_e/N$ then emerges in a phase. This phase correlates the value of $k_e$ with subsets of the spectrum that are singled out through a GSO projection induced by the SS vector ${\bf e}$, which is imposed by the $\beta_e$ sum. The double sum over $\beta_e$ and $\bar{n}$ results in a model that interpolates in the small radius limit to a non-supersymmetric theory in which ${\bf e}$ becomes just another projection vector \cite{Aaronson:2016kjm}. Indeed, this $D+\nu$ dimensional SUSY to $D+\nu$ dimensional non-SUSY interpolation is just a generalization of the phenomenon first observed in Itoyama and Taylor's interpolation from 10D SUSY $E_8\times E_8$ to 10D non-SUSY $SO(16)\times SO(16)$ \cite{Itoyama:1986ei,Itoyama:1987rc,Blum:1997cs,Blum:1997gw}.

\subsection{One-loop cosmological constant (\texorpdfstring{$\Lambda_1$}{Lambda1}) at large radius}

\label{subsec:generic}

Thus, the effect of a Scherk-Schwarz action ultimately results in a breaking of SUSY, yielding a nonvanishing partition function. This rather entangled structure makes complete analytic expressions for the resulting one-loop cosmological constant hard to come by; however, one can deduce
an approximation for it that is valid at relatively large radius by neglecting the exponentially suppressed contribution from the modes with non-zero winding. Indeed, it is clear from Eq.~\eqref{eq:shiftenZint} that the contribution to the partition function coming from the $\bar{n}=0$ spectrum would be identical to that of the unbroken theory were it not for the shift ${\bf e\cdot Q}$ in the KK number. Upon Poisson resumming the $\calZ_{D+\nu,0,k_e}$ factor, we can therefore deduce a generic expression for the total partition function in terms of the unbroken spectrum. Focussing for concreteness on SS actions that act degenerately on the two $\mathbb{T}_2$ dimensions with a half-integer shift in the KK number, ${\bf e\cdot Q}=0,1/2$, we find the following approximation for the partition function at large radius:
\begin{align}
\mathcal{Z}_{{\bar n}=0} (\tau) ~&=~ 2 \tau_2^{1-\frac{D+\nu}{2}}\, \frac{V_\nu } {{\alpha'}^{\nu/2}}~\sum_{\stackrel{M^2_L,M^2_R}{\ell_1+\ell_2=odd}}(n_{b}^{(M^2_{L},M^2_{R})}-n_{f}^{(M^2_{L},M^2_{R})})~\times \\ 
& \qquad\qquad\qquad\qquad\qquad\qquad\qquad e^{-\frac{\pi}{\tau_{2}}\ell^{i}G_{ij}\ell^{j}}
 e^{-\pi M^{2}\alpha'\tau_{2}}
 e^{i\pi\alpha'(M_{L}^{2}-M_R^{2})\frac{\tau_{1}}{2}}~,
\nonumber 
\end{align}
where $n_{b}^{(M^2_L,M^2_R)}$ and $n_{f}^{(M^2_L,M^2_R)}$ count the numbers of bosonic and fermionic states at left- and right-moving levels $M^2_L,M^2_R$ that remain completely unshifted by the action of the Scherk-Schwarz (i.e. that have ${\bf e\cdot Q}=0$), and where $M^2=(M_{L}^2+M_R^2)/2$. 

This expression for the dominant contribution is very general. It says that, in the canonical case in which the SS action results in KK shifts of integer or half-integer values, to determine the partition function one need simply count at any excitation level the net number of states $n_{b}^{(M_L^2,M_R^2)}-n_{f}^{(M_L^2,M_R^2)}$ that remain {\it unshifted}, and double their contribution to find the overall contribution to $\cal{Z}$ from that level.
 (More general SS actions would not add much morally, but the shifted states would be distributed over more values of ${\bf Q\cdot e}$, which would result in much more cumbersome expressions with factors of $\cos (2\pi (\ell_1+\ell_2) {\bf Q\cdot e})$)

The corresponding contribution to the cosmological constant $\Lambda_1$ then follows from the usual integral
\begin{equation}
\Lambda_1(G)~=~-\frac{{\cal M}^{D}}{2}\int_{\mathcal{F}}\frac{d^{2}\tau}{\tau_{2}^{2}}\mathcal{Z}(\tau)~,
\label{eq:1looplam}
\end{equation}
where ${\cal M}^2 = 1/4\pi^2 \alpha'$. 
This integral can be further approximated by also neglecting the exponentially suppressed unphysical mode contribution\footnote{One could alternatively invoke Eq.~\eqref{eq:lamlam}. 
This equation is a valid representation of Eq.~\eqref{eq:1looplam} for any finite $\Lambda_1$ thanks to modular invariance. In the context of string theory it was derived in Ref.~\cite{Dienes:1994jt,Dienes:1994np,Dienes:1995gp,Dienes:1995pm}, and in the general study of integrals of automorphic functions was proven by Rankin and Selberg in Refs.~\cite{rankin1a,rankin1b,selberg1} and was later given an important extension to functions not of rapid decay by Zagier \cite{rankin1a,rankin1b,selberg1,zag}. As we shall ultimately require also the two-loop cosmological constant, it is sufficient for this discussion to consider the one-loop integral in its direct form.}.
Indeed in order to evaluate it we can invoke the exponent 
in $e^{-\frac{\pi}{\tau_{2}}\ell^{i}G_{ij}\ell^{j}}$, which tells us that the dominant contribution comes from the region $\tau_2 ~\gtrsim~ V_\nu^{1/\nu} /\sqrt{\alpha'} $, and that contributions near the $\tau_2\sim 1$ boundary are exponentially suppressed. Hence we can approximate the integral by extending the lower boundary of the fundamental domain ${\cal F}$ down to $\tau_2=0$, and instead performing the integral over the critical strip, ${\cal S} = \{ -1/2 \leq \tau_1 \leq 1/2, \tau_2 \geq 0 \}$. The resulting integral gets nonzero contributions only from the physical level-matched states  with $M_{L}^2=M_{R}^2\equiv M^2$, hence\footnote{Note that the series is uniformly convergent in the case of Scherk-Schwarz breaking, due to the parametrically small breaking of supersymmetry, so we may swap the order of integration and summation.}
\begin{equation}
\Lambda_1(G)~\approx~{{\cal M}^{D}} \frac{V_\nu } {{\alpha'}^{\nu/2}}~ 
\sum_{\stackrel{M^2}{\ell_1+\ell_2=odd}}
 (n_{f}^{(M^2)}-n_{b}^{(M^2)}) \ 
\int_0^\infty \frac{d\tau_2}{\tau_{2}^{1+\frac{D+\nu}{2}}}
 e^{-\frac{\pi}{\tau_{2}}\ell^{i}G_{ij}\ell^{j}}
 e^{-\pi M^{2}\alpha'\tau_{2}}~.
\end{equation}
These integrals can be easily evaluated for both the level zero and massive case. The level zero modes gives a familiar Casimir energy contribution:
\begin{equation}
\label{eq:Lam1Gk=0}
\Lambda_1^{(M^2=0)}(G) ~=~\frac{\left(n_{f}^{(0)}-n_{b}^{(0)}\right) } {(2\pi R)^D } \
 \Gamma\left(\frac{D+\nu}{2}\right) 
 \sum_{ \ell_1+\ell_2=odd } 
 \left( {\pi \ell^i \tilde{G}_{ij} \ell^j}\right) ^{-\frac{D+\nu}{2}} ~.
 \end{equation}
Since in this work we are focusing on theories that have accidental level zero cancellation, $(n_{f}^{(0)}-n_{b}^{(0)}) =0$, we are actually more interested in the massive mode contribution. To write these let us define a rescaled mass, which gives the number of Compton wavelengths within the compactification size $2\pi R$ (and hence essentially captures how far the state is from massless propagation):
\begin{align}
\tilde{M}_{\vec{\ell}} & ~=~\frac{M}{\mathcal{M}}\sqrt{\ell^i G_{ij}\ell^j }\nonumber \\
& ~=~ 2\pi R M \, \sqrt{\ell^i \tilde{G}_{ij}\ell^j }~.
\end{align}
In terms of these rescaled masses the contributions from individual states is  
\begin{equation}
\Lambda_{1}^{(M^2)}~=~
\frac{\left(n_{f}^{(M^2)}-n_{b}^{(M^2)}\right) } {(2\pi R)^D } 
~ 2^{1-\frac{D+\nu}{2}}\sum_{\ell_1+\ell_2={\rm odd}}
\tilde{M}_{\vec{\ell}}^{\frac{D+\nu}{2}}K_{\frac{D+\nu}{2}}({\tilde M}_{\vec{\ell}})~
 \left( {\pi \ell^i \tilde{G}_{ij} \ell^j}\right) ^{-\frac{D+\nu}{2}}~. \label{eq:genk} 
\end{equation}
Since $\lim_{x\rightarrow 0}\left( 2^{1-s} x^s K_s(x) \right) = \Gamma(s)$, Eq.~\eqref{eq:genk} is in fact valid for all $k$. 
By contrast for large masses the sum is exponentially suppressed by the aforementioned Yukawa interaction factor of a massive state, $\exp(-2\pi R M)$, and we can make this explicit by replacing the Bessel function
with its exponential approximation (a.k.a. the saddle-point approximation
to the integral), $K_{s}(x)\rightarrow\sqrt{\frac{\pi}{2x}}e^{-x}$,
so that  Eq.~\eqref{eq:genk} becomes 
\begin{equation}
\Lambda_{1}^{(M^2)}~\stackrel{MR\,\gg\, 1}{\longrightarrow} ~
\frac{\left(n_{f}^{(M^2)}-n_{b}^{(M^2)}\right) } {(2\pi R)^D } 
~\sum_{\ell_1+\ell_2={\rm odd}}
\left(\frac{\tilde{M}_{\vec{\ell}}}{2\pi
}\right)^{\frac{D+\nu-1}{2}}e^{{-\tilde M}_{\vec{\ell}}}~
 \left( {\ell^i \tilde{G}_{ij} \ell^j}\right) ^{-\frac{D+\nu}{2}}~. \label{eq:genk0} 
\end{equation}
Note that in these leading large volume expressions all string scales have cancelled out, so that in effect they describe an extra-dimensional field theory Casimir energy, even for the massive modes.

\subsection{\texorpdfstring{$\Lambda_1$ at large radius on a $\mathbb{T}_2$ torus}{Lambda1 at large radius on a T2 torus}}

\label{subsec:PS}
The specific model we shall focus on for this study is the semi--realistic Bose--Fermi degenerate Pati-Salam model discussed in Ref.~\cite{Abel:2015oxa}. It is based on a large compactification factor on a $\mathbb{T}_2$ torus. For our study, we wish to stabilize all the large moduli and to avoid presuming any kind of symmetric configuration, for example one with degenerate moduli. Therefore, our main task for the one-loop part of the discussion will be to generalize the analysis of Ref.~\cite{Abel:2015oxa} to the case of an
arbitrary $\mathbb{T}_2$ torus.

As a warm-up, let us adapt the large volume approximations of the previous subsection to the case of theories that are compactified on such an arbitrary $\mathbb{T}_{2}$ torus, with dimensionless metric and $B$-field $G_{ij}$ and $B_{ij}$, where
\begin{align}
G_{ij} & \,=\,\frac{T_{2}}{U_{2}}\left(\begin{array}{cc}
1 & U_{1}\\
U_{1} & |U|^{2}
\end{array}\right)\,,
~B_{ij} \,=\, {T_{1}}\left(\begin{array}{cc}
0 & 1\\
-1 & 0
\end{array}\right)\,,
\end{align}
and where 
\begin{eqnarray}
iU & = & U_{1}+iU_{2}\nonumber \\
iT & = & T_{1}+iT_{2}~.
\end{eqnarray}
\begin{align}
\tilde{G}_{ij} & \,=\,\frac{1}{U_{2}}\left(\begin{array}{cc}
1 & U_{1}\\
U_{1} & |U|^{2}
\end{array}\right)\,.
\end{align}
Eq.~\eqref{eq:Lam1Gk=0} yields the massless large volume contribution, 
\begin{equation}
\Lambda_{1}^{(M^2=0)}~=~2\frac{\mathcal{M}^{4}}{T_{2}^{2}}(n_{f}^{(0)}-n_{b}^{(0)}) \,\tilde{E}(3,U)~,
\end{equation}
where the restricted Eisenstein sum is 
\begin{equation}
\tilde{E}(s,U)~=~\pi^{-s}\sum_{\ell_{1}+\ell_{2}=odd}\left(\gamma\cdot U_{2}\right)^{s},
\end{equation}
where
\begin{equation}
\gamma\cdot U_{2}~=~\frac{U_{2}}{|\ell_{1}+U\ell_{2}|^{2}}~.
\end{equation}
This also has a realization in terms of the standard non-holomorphic Eisenstein series, which has an unrestricted sum, $E(s,U)$.
Explicitly, note that $2^{-3}E(3,2U)$ is equivalent to restricting to $\ell_{2}=even$ in the sum, while $2^{-3}E(3,U/2)$ is equivalent to $\ell_{1}=even$
 in the sum, while $2^{-6}E(3,U)$ is equivalent to the sum restricted to $\ell_1$ and $\ell_2$ both even: therefore the $\ell_{1}+\ell_{2}=odd$ sum can be
written 
\begin{align}
\sum_{\ell_{1}+\ell_{2}=odd} & ~=~
\mbox{\small $
\left[
\sum_{\ell_{1}\in 2\mathbb{Z},\ell_{2}}-\sum_{\ell_{1}\in 2\mathbb{Z},\ell_{2}\in 2\mathbb{Z}}\right]$} 
+ \mbox{\small $
\left[
\sum_{\ell_{1},\ell_{2}\in 2\mathbb{Z}}-\sum_{\ell_{1}\in 2\mathbb{Z},\ell_{2}\in 2\mathbb{Z}}\right]$}~, \nonumber \\
& \qquad\implies~~\nonumber \\
\tilde{E}(3,U)& ~=~ 
2^{-3}E(3,U/2)+2^{-3}E(3,2U) -2^{-5}E(3,U) \,\,\,.
\end{align}
For the contributions of the massive states, the rescaled mass becomes
\begin{align}
\tilde{M}_{\vec{\ell}} & ~=~\frac{M}{\mathcal{M}}\sqrt{\frac{ T_{2}}{\gamma\cdot U_{2}}}~.
\end{align}
Thus from Eq.~\eqref{eq:genk0}  the contribution to the cosmological constant at large volume is approximately 
\begin{equation}
\Lambda^{(M^2)}_1~\approx~\frac{\mathcal{M}^{4}}{T_{2}^{2}}\sum_{\ell_1+\ell_2=odd}(n_{f}^{(M^2)}-n_{b}^{(M^2)})\left( \frac{\tilde{M}_{\vec{\ell}}}{2\pi}\right)^{\frac{5}{2}}e^{-\tilde{M}_{\vec{\ell}}}\,(\gamma\cdot U)^{3}.
\end{equation}

\subsection{\texorpdfstring{$\Lambda_1$ at general radius in the $\mathbb{T}_2$ Pati-Salam model}{Lambda1 at general radius in the T2 Pati-Salam model}}

We saw in Eq.~\eqref{eq:genk} that the first subleading contribution to the comsological constant as $R$ shrinks is given by an exponentially suppressed Casimir energy, which has no string scale in the final result except for the internal contributions to the mass of the state itself. This is as it should be because we neglected both unphysical modes and winding modes. What remains therefore to dominate the large radius contribution from a state is either a massless Casimir energy, or a Casimir energy that is given by the Yukawa propagation governed by its mass. Thus, it was possible to make the large radius expressions very general and to derive model agnostic expressions in terms of the light states, to which the compactification lattice simply adds an inert KK tower. 
This is no longer the case when we go to a small compactification radius where the above approximation is expected to break down and where $\alpha'$ dependent contributions can become dominant. What then is the general shape we expect for $\Lambda_1$ over the entire domain compactification scale $R = \sqrt{\alpha' T_2}$? 

\begin{figure}
\centering
\includegraphics[keepaspectratio, width=0.8\textwidth]{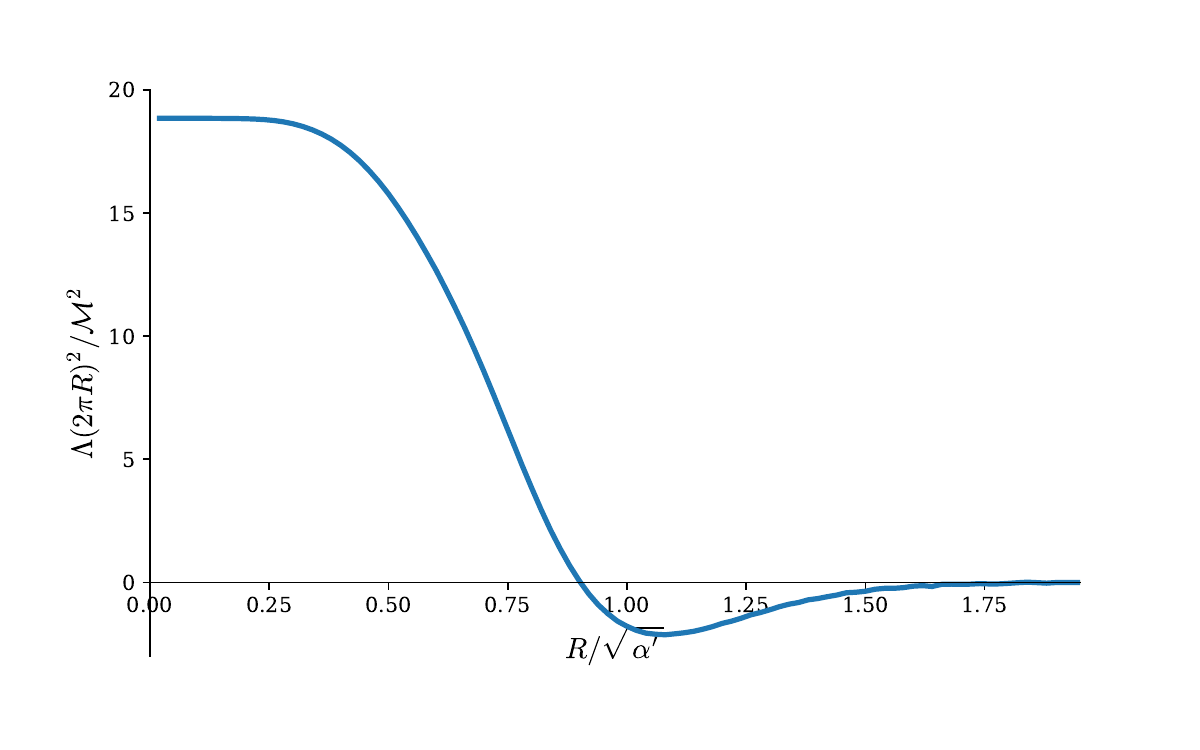}
\caption{Potential (divided by the $R\rightarrow 0$ compactification volume, $V_\nu(R=0)\equiv R^{-2}$) evaluated numerically for the $n_b^{(0)} = n_b^{(0)}$  Pati-Salam model of Ref.~\cite{Abel:2015oxa}, which has $D=4,~ \nu = 2$. Here the compactification is on a square $\mathbb{T}_2$ with degenerate radii $R$ (such that $T_2 = R^2/\alpha'$). This form of potential with positive $\Lambda_1$ at the origin and a dip at $R\sim 1$ is typical of the models of interest. \label{fig:plotscaled} }
\end{figure}

The main additional piece of information we have for these models is that they generally lie on an interpolation between a $D+\nu$ SUSY model at $R\rightarrow\infty$ and a $D+\nu$ non-SUSY model at $R=0$. (When $\nu>1$ interpolation between two supersymmetric $D+\nu$ models is also possible with, as explained in Ref.~\cite{Aaronson:2016kjm}, differently assigned gravitinos at the endpoints; these models would fall into the class considered in Refs.~\cite{Kounnas:2016gmz,Florakis:2021bws}.) Now, we know that the completely non-SUSY model at $R=0$ has a large $D$ dimensional cosmological constant that is proportional to $R^{-\nu}$ (equivalently, there is a constant $D+\nu$ dimensional vacuum energy at $R=0$, where we recall that the compactification volume at $R\rightarrow 0$ goes as $R^{-\nu}$). There is no theorem about what the sign of this cosmological constant should be, but it seems to be generally true that SS models with $n_b^{(0)}=n_f^{(0)}$ only interpolate to either dual SUSY theories or non-SUSY theories with positive cosmological constant. Therefore, if we conjecture that this is always the case, and bearing in mind that $\Lambda_1$ must be negative for the mechanism to work, then for the models of interest we expect to find a dip close to $R\gtrsim 1$ driven by the KK towers of the lowest lying physical states. Continuing to $R=0$, the cosmological constant must then rise again to a positive constant value once the winding modes take over.  We show an example of the overall behaviour in Fig.~\ref{fig:plotscaled}, where we plot the potential of the Bose--Fermi degenerate Pati-Salam model discussed in Ref.~\cite{Abel:2015oxa}, evaluated numerically {\it using the full partition function}.  

As mentioned, Fig.~\ref{fig:plotscaled} and the matching figure in Ref.~\cite{Abel:2015oxa} were produced using the full expression
for the Scherk-Schwarz shifted partition function in Eq.~\eqref{eq:partitionfuncgen}, using the choice of basis vectors and GSO projections for the Pati-Salam model of Ref.~\cite{Abel:2015oxa},
with the assumption of degenerate radii $R_1=R_2=R$. The procedure for evaluating the one-loop potential at all radii to produce the plot in Fig.~\ref{fig:plotscaled} is described in detail Ref.~\cite{Aaronson:2016kjm}.
 That is, in this $N=4$ model there are 16 different possible values of $\bar{n},k_e$, and we determine the $q$-expansion of $\mathcal{Z}_{D+\nu, \, \bar{n}, {k}_e} $ and $\mathcal{Z}_{\text{int},\bar{n} ,{k}_{e}}$ for each. The integrals of all the resulting individual $q$-expansion terms over the fundamental domain in Eq.\eqref{eq:1looplam} can first be  evaluated numerically before being summed, at each given value of $U_2$ and $T_2$ (with for more general radii, $R_1 \equiv \sqrt{\alpha'T_2U_2}$ and $R_2 \equiv \sqrt{\alpha'T_2/U_2}$).

It should be noted that there is a plethora of similar models available which could be considered for this purpose, for example, in Refs.~\cite{Florakis:2021bws,Florakis:2022avh,Faraggi:2020wld}.
Interestingly, such models can already show a dip even in $\Lambda_1$, but in the present case this dip actually turns out to be a saddle once more moduli are included. To see this larger structure, in the present work we wish to be more general and will consider rectangular configurations that still have $U_1=0$, but that have arbitrary $U_2$ and $T_2$ values. Thus, the moduli space we consider is two-dimensional.

This complete treatment will at large radius yield a $\Lambda_1$ that asymptotes to the leading massive Casimir contribution of Subsection~\ref{subsec:PS}, while at small radius it will asymptote to the leading massless Casimir contribution of a dual $R\rightarrow 0$ theory. Ultimately in order for our two-loop approximation (to be discussed in a moment) to be valid, we shall need to form a minimum in the asymptotic Casimir regime, and away from $R=1$. That is, a true minimum will form somewhere up the slope of Fig.~\ref{fig:plotscaled}, but away from $R=1$.

\subsection{Estimating two-loop contributions}

\label{subsec:2loop}

We now proceed to the two-loop cosmological constant terms $\Lambda_2$. These terms are naturally much more complicated; however, as they are not required to cancel, and as we are anticipating minima at relatively large volume, we can consider them at the level of the leading Casimir energy, but retaining the full dependence on $U$. 

The two-loop cosmological constant for such theories was discussed in Ref.~\cite{Abel:2017rch}. We can adopt the contribution from the untwisted+twisted sunset diagrams as being representative:
\begin{align}\label{eq:two-loop_bare}
\Lambda^{(0)}_2~&\approx~\frac{\pi^{-2}}{(2\pi R)^D}(n_{b}^{\rm (tw)}+n_{f}^{\rm (tw)})(n_{b}^{\rm (un)}-n_{f}^{\rm (un)})~\times \nonumber \\
&\qquad \qquad \left( (3+2\log(N/\pi^2)) \tilde{E}(3,U) + \tilde{E}'(3,U)  \right)~,
\end{align}
where $\tilde{E}'(s,U)\equiv \partial_s\tilde{E}(s,U)$, and where the superscripts $`{\rm tw}$' and $`{\rm un}$' indicate that the count is over massless twisted and untwisted degrees of freedom respectively. In the above estimate, the parameter $N$ is a regulator (which goes to infinity in the unregulated theory) required to get a finite answer in the presence of uncancelled divergences. Physically such divergences may be interpreted as a logarithmic running for the one-loop gauge coupling. In other words, in the field theory there would be a counter term that removes it, while the full string theory expression is finite but difficult to compute. In the present case, the natural assumption is that this term is regulated when the theory is stabilized by physical masses: thus one expects a logarithmic dependence on the typical mass scales of the theory such that $N\sim T_2$. This is the point of view we take in the following. To summarize these effects, we parametrize the two-loop contribution as
\begin{eqaed}\label{eq:two-loop_parametrization}
    \Lambda_2^{(0)} ~\approx~ \frac{a}{T_2^2} \left( (b + c \, \log T_2 ) \, \tilde{E}(3,U)  + \tilde{E}'(3,U) \right) \, .
\end{eqaed}
In the specific model we used, taken from \cite{Abel:2015oxa, Abel:2017rch}, there are 2448 massless bosons (and thus fermions). Estimating the prefactor in \cref{eq:two-loop_bare} from the more precise expressions found in Ref.~\cite{Abel:2017rch} requires knowledge of the two-loop degeneracies and tree-level Yukawa couplings. Estimating the former by the squares of the number of massless bosons/fermions and the latter by $\sqrt{2} \, g_\text{YM} \approx \sqrt{2}$ suggests the choice 
\begin{equation}
a ~ = ~ \frac{(\sqrt{2} \, 4896)^2}{\pi^5} ~\approx ~ 10^5~.
\label{eq:a-choice}
\end{equation}

\section{Banks-Zaks Stabilization}\label{sec:banks-zaks}

\subsection{Schematics in general dimensions}

Having gathered the necessary ingredients, let us now return to the Banks-Zaks-like stabilization mechanism proposed in Section~\ref{sec:introduction}, now repeated for $d<10$ total dimensions for completeness. In this fashion, we can take into account more conveniently settings in which an internal sector is frozen, such as in the asymmetric orbifolds of \cite{Baykara:2024tjr, Baykara:2024vss, Angelantonj:2024jtu,Faraggi:2022hut}, whereas $d-D$ dimensions are. In the $d$-dimensional Einstein frame, the effective potential for the dilaton and the two moduli we consider takes the form
\begin{eqaed}\label{eq:einstein-frame_pot}
    V_\text{Einstein} = e^{\frac{2d}{d-2} \phi} \Lambda_1(T,U) + e^{(\frac{2d}{d-2} + 2)\phi} \, \Lambda^{(0)}_2(T,U) \, ,
\end{eqaed}
where again the one-loop term has no net contribution from the massless states, such that it may balance against the two-loop term coming from the massless states. Working with the compactification radius $R$ defined by Eq.~\eqref{eq:vols} and retaining the full $U = i U_2$ dependence, we further approximate the two-loop term according to~\Cref{eq:two-loop_parametrization}, which we denote more succinctly as
\begin{eqaed}\label{eq:2-loop_approx_cc_any_dim}
    \Lambda_2 \approx \zeta(U) \, R^{- D}
\end{eqaed}
where $D$ is the number of extended dimensions. The resulting stabilized values for the Einstein-frame cosmological constant and string coupling are then
\begin{eqaed}\label{eq:final_stabilised_values}
    e^{2\phi} & ~=~ - \, \frac{d}{2(d-1)} \, \frac{\Lambda_1}{\Lambda_2} \, , \\
    \Lambda & ~=~ \frac{d-2}{2(d-1)} \, \Lambda_1 \left( - \, \frac{d}{2(d-1)} \, \frac{\Lambda_1}{\Lambda_2} \right)^{\frac{d}{d-2}} \, ,
\end{eqaed}
which correspond to a (necessarily AdS) minimum if and only if $\Lambda_1 < 0$ and $\Lambda_2 > 0$.

In order to understand the one-loop stringy contribution, it is useful to keep in mind the large radius one-loop expression from Eq.~\eqref{eq:genk0} as a rough indication of the expected behaviour. However, as mentioned, to perform the stabilization we will use the full partition function, which also captures the small-radius behaviour. 

\subsection{Numerical results in four dimensions}

We will now demonstrate that adding the massless two-loop contribution parametrized by~\Cref{eq:two-loop_parametrization} according to~\Cref{eq:einstein-frame_pot}, indeed leads to nontrivial AdS minima as anticipated. 

The minimization of the effective potential of~\Cref{eq:final_stabilised_values} depends on the parameters $a, b, c$ in~\Cref{eq:two-loop_parametrization}. More precisely, the minimum does not depend on $a$ in any respect other than its sign; that is, we require $a>0$. The actual value of $a$ only enters the value of $g_s$ and $\Lambda$ at the minimum, but not the position of the minimum in the (pseudo-)moduli space. Within our approximations, for some ranges of values, there is a minimum, which will always be AdS due to our preceding considerations. Based on our estimates of the two-loop term, it seems reasonable to choose $a \approx 10^5$ as per Eq.~\eqref{eq:a-choice}. Furthermore, in order to perform the minimization explicitly, we restricted the geometric moduli of the torus to their imaginary parts, $T=T_2 \, , \, U=U_2$. As such, the following results should be taken as a proof-of-concept of the stringy Banks-Zaks mechanism.

The most visible minimum that we find appears for $b = 0.15$ and $c = 0$. The resulting cosmological constant is $\Lambda \approx -0.61 \times 10^{-5}$ in string units where $\mathcal{M}=1$, while the moduli attain the values $U_2 \approx 1$ and $T_2 \approx 1.17$. The value of $U_2$ could be predicted by symmetry considerations, a reassuring consistency check. The closed-string coupling is $g_s^2 \approx 2.57 \times 10^{-5}$ at the minimum.

Thus, since $\Lambda$ has units of $\mathcal{M}^4$, assuming that the remaining extra dimensions that we do not consider have a string-size volume $V_4 \,\approx\, \mathcal{M}^{-4}$, we have
\begin{eqaed}\label{eq:mass_parameters}
    M_{\text{KK}} & ~\equiv~ ~(V_2 \, V_4)^{- \frac{1}{6}} ~~~\approx ~\left(\frac{4\pi^2}{T_2}\right)^{\frac{1}{6}} \, \mathcal{M} \, , \\
    M_{\text{Pl}}^2 & ~\equiv~ \frac{2\pi \mathcal{M}^8}{g_s^2} \, V_2 \, V_4 ~\approx~ \frac{T_2}{2\pi g_s^2} \, \mathcal{M}^2 \, ,
\end{eqaed}
and therefore the relevant ratios for the cosmological constant and scale separation in four dimensions are
\begin{eqaed}\label{eq:scale_parameters}
    \frac{\Lambda}{M_{\text{Pl}}^4} & ~\approx~ \frac{4\pi^2 g_s^4}{T_2^2} \, \frac{\Lambda}{\mathcal{M}^4} ~\approx~ - \, 1.31 \times 10^{-13} \, , \\
    \frac{\ell_{4d}^2}{\ell_{\text{KK}}^2} & ~\equiv~ \frac{M_{\text{Pl}}^2 \, M^2_{\text{KK}}}{\abs{\Lambda}} ~~\approx~ \frac{T_2^{\frac{2}{3}}}{(2\pi)^{\frac{1}{3}} \, g_s^2 \, \abs{\Lambda}} ~\approx~ 3.39 \times 10^9 \, .
\end{eqaed}

\begin{figure}[ht!]
    \centering   \includegraphics[trim= 0 0 750 0,clip,width=\textwidth]
{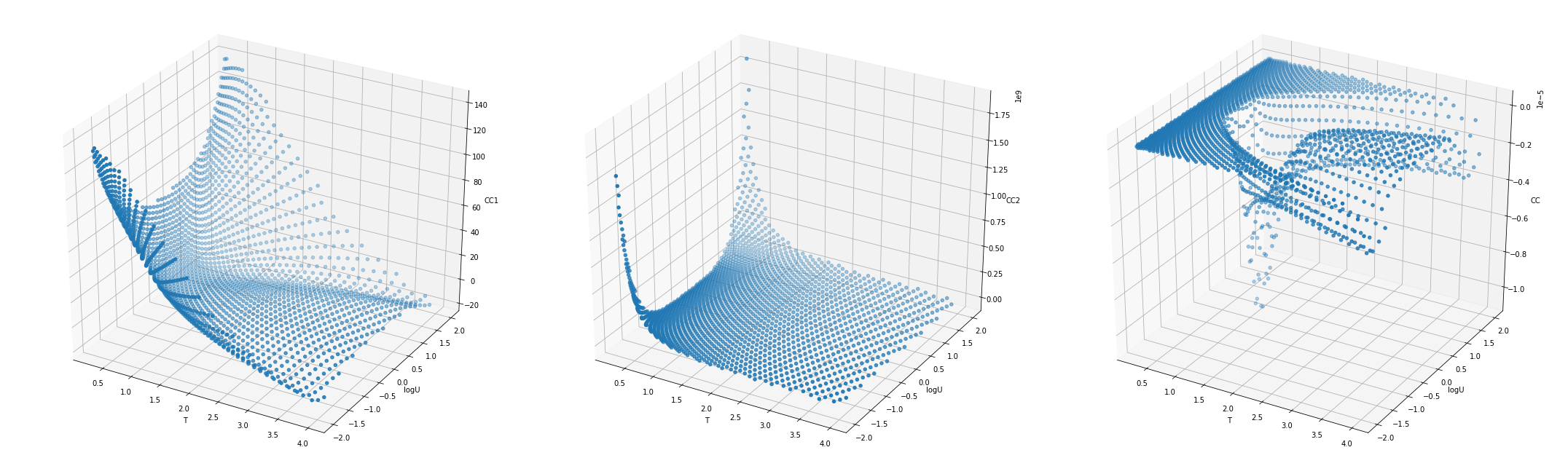}
 \includegraphics[scale=0.32,trim= 1410 0 0 0,clip]
{BS_plot.png}
    \caption{Contributions to the effective potential as functions of $(T_2, \log U_2)$. Top row: the one-loop contribution $\Lambda_1$ and the massless approximated two-loop contribution $\Lambda_2^{(0)}$ of~\Cref{eq:two-loop_parametrization}. Second row: the combined effective potential at the stabilised dilaton value in~\Cref{eq:final_stabilised_values}. The parameters chosen for this plot are $b = 0.15$ and $c = 0$. The value of $a$ only appears as a prefactor which is irrelevant in the minimisation, and thus we set $a=1$ in this plot. The combined effective potential is only defined in the region where $\Lambda_1/\Lambda_2 < 0$, and features a single minimum.}
    \label{fig:BS_plot}
\end{figure}

We stress that these results were obtained by combining the {\it entire} massive one-loop contribution with the (estimated) two-loop contribution in \cref{eq:two-loop_parametrization}. That is, we computed the one-loop contribution numerically in the $(T_2, U_2)$ plane and combined it with Eq.~\cref{eq:two-loop_parametrization}. The total effective potential is therefore only defined in the region $\Lambda_1/\Lambda_2 < 0$, and features a minimum as depicted in \cref{fig:BS_plot}.

\section{Conclusions}\label{sec:conclusions}

    In this paper, we have proposed a general mechanism, which is analogous to the Banks-Zaks mechanism,  that can generate scale-separated vacua. The resulting settings are meta-stable AdS vacua whose scale separation is numerical rather than parametric. This is consistent with various arguments in the literature that prevent the latter \cite{Cribiori:2023gcy, Cribiori:2023ihv, Cribiori:2024jwq, Lust:2022lfc, Bena:2024are} and with the constructions in Refs.~\cite{Demirtas:2021nlu, Demirtas:2021ote, McAllister:2024lnt}.

    Our approach directly addresses the issue of dilaton stabilization. It is known that achieving this at the one-loop level requires the introduction of fluxes \cite{Mourad:2016xbk} and never produces scale separation. In contrast, the Banks-Zaks approach that we have advocated here bypasses the ten-dimensional effective field theory, ending up directly in the lower-dimensional vacuum.

    This paper introduces the concept of Banks-Zaks stabilization, and it remains to be seen how such minima can be successfully incorporated into the phenomenology of realistic models. At the technical level, our computation relied on several estimates of the two-loop contribution to the vacuum energy. In addition, questions remain over the potential non-perturbative decay of these AdS vacua. Since fluxes are not involved, the most natural decay channel to consider would be to bubbles of nothing, not obstructed by supersymmetry or topology \cite{GarciaEtxebarria:2020xsr}. In AdS, the corresponding decay rate was discussed in Ref.~\cite{Dibitetto:2020csn}. This kind of decay is likely to be at best numerically suppressed by the scale separation; however, the type of effect we are proposing would need to be incorporated within a more complete picture, and most likely these estimates would change.

    Our proposal opens a number of avenues for further investigation. To begin with, more precise and systematic computations can be performed starting in nine dimensions rather than directly in four. In this case, no moduli would need to be projected out or neglected, and a systematic study along the lines of Ref.~\cite{Fraiman:2023cpa} could be feasible. It would also be interesting to combine this approach with the recent non-supersymmetric rigid constructions of Refs.~\cite{Baykara:2024tjr, Angelantonj:2024jtu}, in order to arrive at four-dimensional vacua that do not have many moduli that require stabilizing. Our Banks-Zaks stabilization mechanism should be able to operate in any such theories that satisfy the necessary conditions we have outlined here.
    
\section*{Acknowledgements}

    I.B. thanks M. Emelin for discussion.


\printbibliography

\end{document}